\documentclass[%
 reprint,showkeys,
 amsmath,amssymb,
 aps,
 prstab,
]{revtex4-2}

\usepackage{graphicx}
\usepackage{dcolumn}
\usepackage{bm}
\usepackage{hyperref}

\begin{document}

\preprint{APS/123-QED}

\title{\textbf{Bunch-by-Bunch Prediction of Beam Transverse Position, Phase, and Length in a Storage Ring Using Neural Networks}}%

\author{Can Liu}
\affiliation{
 National Synchrotron Radiation Laboratory (NSRL), University of Science and Technology of China, Hefei 230029, China
}

\author{Xing Yang}
\affiliation{
 National Synchrotron Radiation Laboratory (NSRL), University of Science and Technology of China, Hefei 230029, China
}

\author{Youming Deng}
\affiliation{
 National Synchrotron Radiation Laboratory (NSRL), University of Science and Technology of China, Hefei 230029, China
}

\author{Qingqing Duan}
\thanks{Corresponding author. Email: duanqq0701@gmail.com}
\affiliation{
 National Synchrotron Radiation Laboratory (NSRL), University of Science and Technology of China, Hefei 230029, China
}

\author{Yongbin Leng}
\thanks{Corresponding author. Email: lengyb@ustc.edu.cn}
\affiliation{
 National Synchrotron Radiation Laboratory (NSRL), University of Science and Technology of China, Hefei 230029, China
}

\date{December 19, 2025}

\begin{abstract}
Real-time, bunch-by-bunch monitoring of transverse position, longitudinal phase, and bunch length is crucial for beam control in diffraction-limited storage rings, where complex collective dynamics pose unprecedented diagnostic challenges. This study presents a neural network framework that simultaneously predicts these parameters directly from beam position monitor waveforms. The hybrid architecture integrates specialized Multi-Layer Perceptron (MLP), Convolutional Neural Network (CNN), and Long Short-Term Memory with Attention (LSTM-Attention) sub-networks, overcoming key limitations of traditional methods such as serial processing chains and batch-mode operation. Validated on experimental data from the Shanghai Synchrotron Radiation Facility and Hefei Light Source, the model achieves high prediction accuracy with a sub-millisecond latency of 0.042 ms per bunch. This performance demonstrates its potential as a core tool for real-time, multi-parameter diagnostics and active feedback in next-generation light sources.
\end{abstract}


\maketitle

\section{Introduction}
\label{sec:intro}

\subsection{Research Background and Significance}
\label{sec:background}

Synchrotron radiation sources are entering the era of diffraction-limited storage rings (DLSRs), which operate with extremely low emittance ($\sim 100\ \text{pm}\cdot\text{rad}$), high beam current, and in multi-bunch mode ($> 500$ bunches) \cite{tanaka2018}. In next-generation facilities, such as the Hefei Advanced Light Facility (HALF) and the High Energy Photon Source (HEPS), the use of small-aperture vacuum chambers and narrow-gap insertion devices significantly amplifies collective effects (e.g., wakefields, instabilities) and enhances inter-bunch coupling. This results in a complex beam system exhibiting multi-scale, nonlinear, and time-varying dynamic characteristics, which introduces unprecedented challenges for beam diagnostics and control.

Conventional turn-by-turn averaging diagnostic methods are inadequate for understanding and controlling the complex collective phenomena in DLSRs due to the masked individual bunch dynamics\cite{chen2014}. Moreover, currently deployed bunch-by-bunch processing systems (e.g., HOTCAP) are also limited, as exemplified by the cross-correlation-based architecture of HOTCAP itself, which relies on a predefined global response function assuming uniform bunch length\cite{xu2021}. This assumption is often invalid under real multi-bunch operation, making direct bunch length measurement unfeasible for such a system. Furthermore, its sequential processing flow (phase $\rightarrow$ amplitude $\rightarrow$ position) leads to error propagation and accumulation, while its batch-processing nature results in high latency (on the order of seconds), rendering it unsuitable for the real-time diagnostics and feedback required for stable DLSR operation\cite{yang2024}. Consequently, the development of new real-time bunch-by-bunch diagnostic techniques capable of resolving multi-parameter dynamics has become a pressing need for the commissioning, operation, and physics research of DLSRs\cite{yang2015}. This technology provides physicists with a direct ``macro-particle'' observational perspective, offers operation staff critical tools for real-time monitoring and optimization, and represents a key step in the ongoing paradigm shift in beam diagnostics\cite{deng2024}.

In response to these limitations, deep learning, with its powerful end-to-end feature extraction and parallel processing capabilities, offers a new paradigm for establishing direct mapping from raw signals to multi-dimensional beam parameters (bunch length, phase, position)\cite{bengio2009learning}. For instance, neural networks have been applied to achieve millisecond-level real-time identification of beam loss sources and high-precision prediction of longitudinal phase space distributions from non-invasive beam parameters. Fermi Lab successfully implemented real-time identification of beam loss sources in the Main Injector and Recycler Ring using a UNet architecture, achieving 85\%–94\% classification accuracy with only 1.7 milliseconds of delay\cite{hazelwood2023}; Additionally, SLAC employed a dual-stream multilayer perceptron to achieve high-precision prediction of longitudinal phase space distributions based on non-destructive beam parameters, controlling errors in peak current and bunch length predictions to within 5\% and 3\%, respectively\cite{emma2018}. This significant progress demonstrates that research is shifting toward data-driven approaches to overcome the assumptions and limitations of traditional physical models by fully leveraging the physical information embedded in large-scale beam diagnostic data\cite{fol2018,ji2024,zmmr-ry9h}.

Building on this paradigm and directly targeting the specific needs of DLSRs, this study proposes a hybrid neural network architecture that integrates specialized feature learning, parallelized multi-parameter prediction, and ultra-low-latency processing. This integration is realized through three dedicated prediction branches within a unified framework. Our architecture enables the end-to-end joint prediction of bunch length, phase, and transverse position through the collaborative design of shared feature extraction layers and dedicated prediction branches. Specifically, the bunch length prediction branch utilizes multi-scale dilated convolutions to effectively capture waveform time-domain features. The phase prediction network adds a BiLSTM-Attention mechanism to accurately model long-range dependencies. The position prediction module employs residual connections to enhance feature reuse capability.

\subsection{Introduction to the Source}
\label{sec:source}

To develop and validate the proposed bunch-by-bunch diagnostic method, experimental data were acquired from two representative synchrotron radiation facilities in China: the Shanghai Synchrotron Radiation Facility (SSRF) and the Hefei Light Source II (HLS-II).

The Shanghai Synchrotron Radiation Facility (SSRF) is a flagship third-generation light source. Its storage ring is designed for low emittance and stable multi-bunch operation, providing an ideal platform for obtaining high-quality, steady-state beam data. Under top-up injection mode, SSRF achieves exceptional beam stability, with orbit jitter maintained at the sub-micrometer level through advanced feedback systems. This high stability means that data collected from SSRF primarily reflect the intrinsic behavior of the beam under well-controlled conditions, with negligible instabilities, making it suitable for benchmarking diagnostic algorithms in a "quiet" environment \cite{yin2016}.

In contrast, the Hefei Light Source II (HLS-II), as an upgraded second-generation facility, exhibits distinct beam dynamics characteristics. It operates with a different RF frequency and bunch spacing, and more importantly, its beam is subject to noticeable longitudinal coupled-bunch oscillations during routine operation. Data from HLS-II thus inherently contain richer dynamic information, including the effects of instabilities. Furthermore, the beam-based alignment (BBA) system and closed-orbit feedback system implemented at HLS-II ensure micron-level beam orbit stability, establishing a reliable experimental foundation for studying bunch-by-bunch parameter dynamics \cite{wu2017}. Using HLS-II data allows for testing the robustness and generalization capability of diagnostic models under more challenging, "noisy" conditions.

The strategic selection of these two sources ensures that the developed neural network framework is trained and validated on datasets covering a wide spectrum of operational scenarios---from the ultra-stable conditions of a mature third-generation light source to the more dynamic environment of a second-generation ring with active instabilities. This approach enhances the model's practicality and preparedness for future applications in next-generation facilities like HALF and HEPS, where both extreme stability and complex instability management will be required.

\section{Dataset Composition}
\label{sec:dataset}

Experimental data were acquired from broadband beam position monitors (BPM) at the Shanghai Light Source (SSRF) and the Hefei Light Source (HLS-II), with both facilities operating at a sampling frequency of 16 GHz. The datasets represent two distinct operational regimes: SSRF data were collected under empty ring injection mode with 9 bunches over 7736 turns, while HLS-II data correspond to multi-bunch steady-state operation with 35 bunches across 56673 turns. Signals from four BPM electrodes were synchronously recorded and stored in MATLAB format.

The fundamental difference in RF frequencies between the two storage rings results in significantly different waveform sampling characteristics. This inherent disparity necessitates distinct data slicing strategies for each facility and introduces a systematic phase sampling error that accumulates both across bunches and over successive turns. To address this critical issue, we introduced a Tshift compensation term during dataset construction. This compensation mechanism enables effective correction of the phase drift arising from discrete sampling of continuous bunch signals, thereby ensuring robust model performance across both operational regimes.

The following sections of this chapter will describe how the collected raw data is structured into the dataset required for our experimental training.

\subsection{Raw Data Slicing}
\label{sec:raw_slicing}

Data preprocessing begins with normalizing the four-channel electrode signals to eliminate ADC range differences. The baseline region of the signal is detected using a sliding window method. When the peak-to-peak difference within consecutive windows is less than 0.1 times the signal amplitude, the segment is identified as a baseline, and its mean value is taken as the DC offset for subtraction. Bunch localization employs a dynamic threshold method: a primary threshold (0.25 times the amplitude) identifies the rising/falling edges of the bunch, followed by a search within a local window (20 sampling points) for the voltage minimum point, which is marked as the bunch center position.

For multi-bunch operation mode, the algorithm calculates the theoretical bunch spacing based on the RF frequency:

\begin{equation}
N = \frac{f_{s}}{f_{RF}}
\label{eq:bunch_spacing}
\end{equation}

where the sampling frequency $f_{s}$ is 16 GHz; the SSRF RF frequency is 499.654 MHz, resulting in 32.02 sampling points; the HLS-II RF frequency is 204.027 MHz, resulting in 78.42 sampling points. After rounding, these become 32 and 78 points, respectively. The sampling phase error introduced by this rounding operation is the direct reason for implementing the Tshift compensation mechanism later.

After calculating the theoretical bunch spacing, the slicing window is dynamically adjusted based on the actual bunch positions. The extracted four-channel voltage waveform segments for each bunch are then normalized to eliminate amplitude variations caused by differences in beam current or electronic gain. This process results in standardized waveform inputs for subsequent processing. The integrity of these preprocessed waveforms forms the basis for subsequent analysis. Recent advancements in signal reconstruction with phase compensation further underscore the importance of high-fidelity raw signal acquisition for bunch-by-bunch diagnostics \cite{deng2024b}. 

\begin{figure}[h]
\centering
\includegraphics[width=0.8\linewidth]{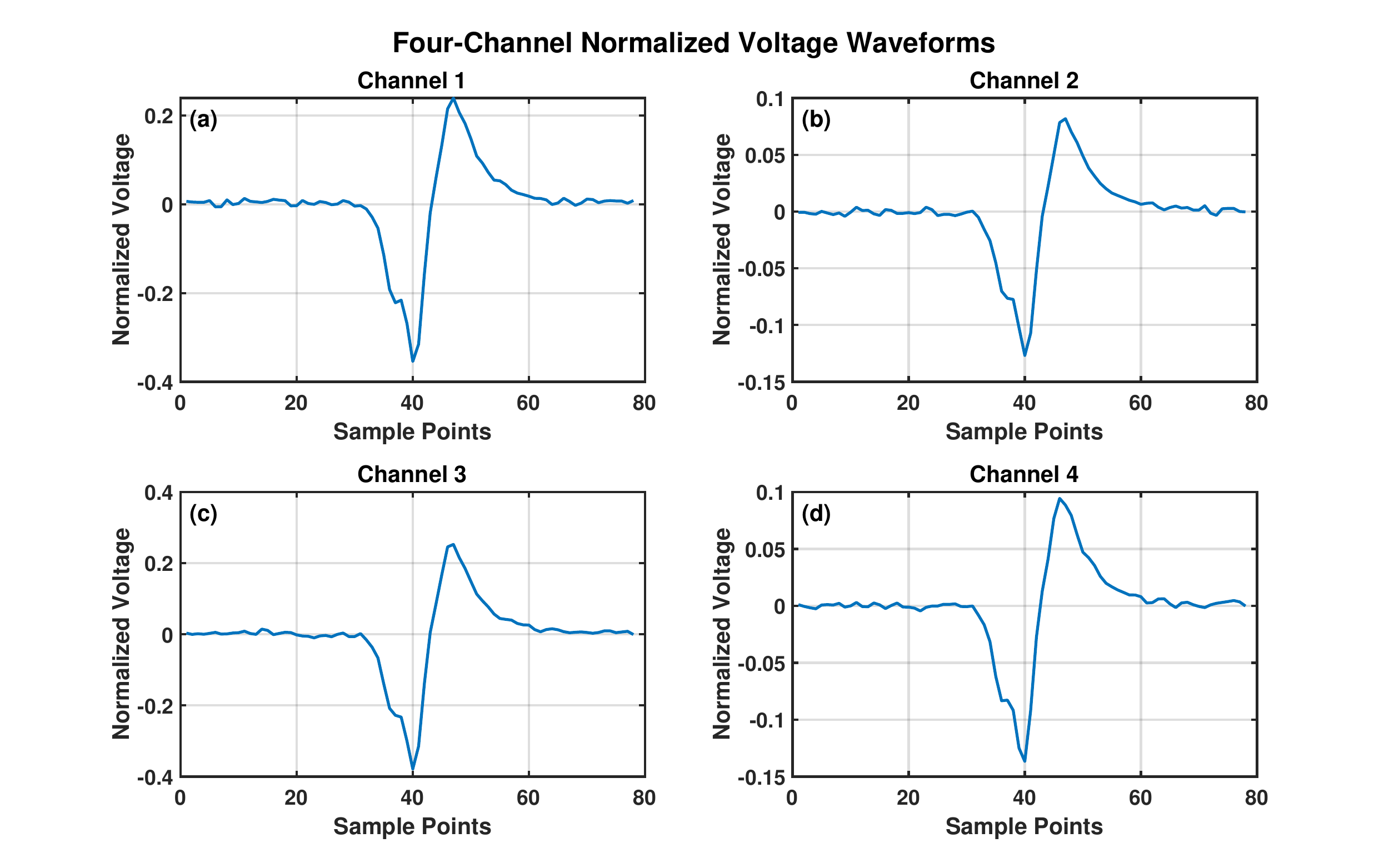}
\caption{Normalized BPM voltage waveforms of the first bunch in the first revolution at HLS-II.}
\label{fig:bpm_waveforms}
\end{figure}

\subsection{Calculating Theoretical 3D Position}
\label{sec:3d_position}

The ground truth calculation for the 3D position is performed by the HOTCAP program \cite{xu2021b}. Its core algorithm uses cross-correlation to reconstruct the beam's 3D position: First, for the four BPM electrode signals of each sliced bunch, the program performs waveform matching based on a pre-constructed lookup table (LUT) obtained from multi-period averaging. By calculating the cross-correlation function between the sliced signal and the LUT template, the optimal phase offset is determined. After phase alignment, the amplitude is extracted by comparing with the LUT template, and the transverse position is calculated using the difference-over-sum formula:

\begin{align}
x &= k_{x} \cdot \frac{{amp}_{A} - {amp}_{B} - {amp}_{C} + {amp}_{D}}{{amp}_{A} + {amp}_{B} + {amp}_{C} + {amp}_{D}} \label{eq:x_position} \\
y &= k_{y} \cdot \frac{{amp}_{A} + {amp}_{B} - {amp}_{C} - {amp}_{D}}{{amp}_{A} + {amp}_{B} + {amp}_{C} + {amp}_{D}} \label{eq:y_position}
\end{align}

where $k_{x}$, $k_{y}$ are the BPM position sensitivity coefficients, which are calibration constants, and amp represents the amplitude of each electrode signal. This established method relies on pre-calibrated coefficients. Meanwhile, alternative techniques pursuing direct, calibration-free position measurement have been explored, such as using time-of-arrival differences at multiple electrodes \cite{yang2024b}.

\subsection{Calculating Theoretical Bunch Length}
\label{sec:bunch_length}

The ground truth calculation for the bunch length is completed using a joint time-frequency domain analysis technique \cite{wang2024}. This method converts the time-domain waveform of a single bunch into the frequency domain via Fast Fourier Transform (FFT). In the frequency domain analysis, for a Gaussian-distributed bunch, there is a clear mathematical relationship between the power spectral distribution of the beam signal and the bunch length:

\begin{equation}
V(f) = I(f) \cdot R_{\text{Button}}(f) = Q \cdot \exp\left( - \frac{\omega^{2}}{2\sigma_{f}^{2}} \right) \cdot R_{\text{Button}}(f)
\label{eq:bunch_length_freq}
\end{equation}

where $V(f)$ is the frequency domain distribution of the measured signal, $I(f)$ is the frequency domain distribution of the current signal, $\sigma_{f}$ is the reciprocal of the bunch length in the time domain, and $R_{\text{Button}}(f)$ is the frequency-domain transfer impedance determined by calibration experiments. The measured voltage spectrum $V(f)$ is divided by the frequency-domain transfer impedance to obtain the current spectrum $I(f)$, and then Gaussian fitting is performed on the current spectrum to extract the bunch length parameter. This methodology for bunch-by-bunch length extraction is well-established and has been implemented in dedicated diagnostic systems \cite{wang2025}.

\subsection{Tshift Compensation Input}
\label{sec:tshift}

The introduction of the Tshift compensation term is essentially due to the non-integer multiple relationship between the beam signal period (determined by the RF frequency) and the sampling period (determined by the ADC sampling rate). This non-integer multiple relationship causes the initial sampling phase of each bunch to shift accumulatively over different turns, resulting in systematic differences in the initial phase of different bunches. Without compensation, this phase misalignment would introduce periodic errors in subsequent processing, affecting the prediction accuracy of phase and position parameters.

Due to the non-integer multiple relationship between the RF frequency and the sampling rate, the theoretical bunch spacing has a fractional sampling point (SSRF: 32.02 points, HLS-II: 78.42 points). The rounding operation introduces a periodic phase accumulation error. To address this, a Tshift compensation term is constructed for dynamic updating and compensation. Its calculation is:

\begin{equation}
Tshift = \text{frac}(N) \cdot n \cdot k
\label{eq:tshift}
\end{equation}

where $\text{frac}(N)$ is the fractional part of the number of sampling points; $n$ is the turn number; $k$ is the filling number of this bunch in the storage ring. This data serves as a supplementary input for the phase prediction part to compensate for the error introduced by the rounding operation.

\subsection{Dataset Division}
\label{sec:dataset_division}

After the aforementioned signal processing and parameter compensation steps, two independent datasets were finally constructed (Table \ref{tab:datasets}).

An equidistant sampling strategy was adopted, extracting 5\% of samples from each dataset as the test set (SSRF: 3,482 samples, HLS-II: 99,178 samples), with the remaining 95\% used as the training set. The sampling interval was 1 out of every 20 samples, ensuring the test set uniformly covered all bunches and operational states.

\begin{table}[h]
\caption{SSRF and HLS-II datasets. 
Data were collected from broadband BPMs at SSRF and HLS-II facilities. 
The SSRF dataset corresponds to empty ring injection mode, while HLS-II data 
represent multi-bunch steady-state operation. The Tshift compensation term 
accounts for phase drift due to non-integer sampling of RF periods. 
SSRF: Shanghai Synchrotron Radiation Facility; HLS-II: Hefei Light Source II.
$\phi$: longitudinal phase; $\sigma$: bunch length.
}
\label{tab:datasets}
\begin{ruledtabular}
\begin{tabular}{lcc}
 & \textbf{SSRF} & \textbf{HLS-II} \\
\hline
Single Sample Input Dim & $4\times32 + 1$ (Tshift) & $4\times78 + 1$ (Tshift) \\
Output Parameters & $x, y, \phi$ & $x, y, \phi, \sigma$ \\
Number of Bunches & 9 & 35 \\
Number of Turns & 7736 & 56673 \\
Total Sample Size & 69624 & 1983555 \\
\end{tabular}
\end{ruledtabular}
\end{table}

\section{Model Preparation}
\label{sec:model}

The machine learning-based beam diagnostic system established in this study adopts a unified data input framework, using the waveform signals from four BPM electrodes and the time compensation Tshift as basic inputs. Specialized neural network modules are designed according to the characteristic differences of different physical quantities. The core design philosophy of the system is to optimize the network structure based on the essential characteristics of each physical quantity. In the following sections detailing each module's data processing workflow, the HLS-II dataset will be used as the primary example for illustration.

\subsection{Prediction Model Module Introduction}
\label{sec:modules}

\subsubsection{Multilayer Perceptron Module}
\label{sec:mlp}

The Multilayer Perceptron (MLP), as a classic neural network structure, is used in this study for prediction tasks that do not rely on time-series characteristics \cite{rumelhart1986}. The MLP consists of fully connected layers, where each neuron is connected to all neurons in the adjacent layer, achieving complex feature transformations through nonlinear activation functions. For transverse position prediction, considering its physical essence is the difference-over-sum calculation of the amplitudes of the four BPM signals, which does not involve the temporal dependence of the waveform, we use an MLP module comprising one input layer, two hidden layers, and one output layer for prediction, employing the ReLU activation function to introduce nonlinearity. The advantage of the MLP lies in its simple and efficient structure, making it particularly suitable for prediction problems with strong global correlations between features. In terms of computational efficiency, the forward propagation of the MLP only involves matrix multiplication and activation function calculations, giving it a clear advantage in inference speed.

\subsubsection{Convolutional Neural Network Module}
\label{sec:cnn}

The Convolutional Neural Network (CNN) is an effective tool for processing local feature extraction \cite{lecun2015, gu2018}. In this study, for predicting bunch length and phase which require consideration of local or global dependencies, we employ a one-dimensional CNN structure to process the beam monitoring data. The input data dimension is [4, 78], where 4 represents the number of BPM electrode channels and 78 is the number of time sampling points. By stacking multiple convolutional layers, the network can automatically learn the spatial correlations between different electrode signals and the temporal features of the signals themselves. Each convolutional layer is followed by a ReLU activation function and a max-pooling operation, gradually compressing the feature dimensions and enhancing translation invariance. Finally, a fixed-dimensional feature representation is obtained through a global average pooling layer.

\subsubsection{LSTM and Attention Mechanism Module}
\label{sec:lstm_attention}

The Long Short-Term Memory (LSTM) network, an improved architecture of Recurrent Neural Networks (RNN), effectively solves the vanishing gradient problem in long sequence modeling through its gating mechanism \cite{sherstinsky2020}. Its unique input gate, forget gate, and output gate structures can selectively update and transmit temporal information, making it particularly suitable for processing beam monitoring data with long-term dependencies.

In this study, the LSTM module receives local time-domain features extracted by a 1D convolutional network as input. The 1D CNN first performs multi-scale feature extraction on the raw BPM signals, capturing local fluctuation patterns in the signal through convolutional kernels; subsequently, the LSTM performs temporal integration of these local features, establishing dependencies across time steps, thereby completely modeling the dynamic evolution process of beam parameters.

The introduction of the attention mechanism further optimizes this architecture. This mechanism calculates the importance weights of features at each time step, enabling the model to adaptively focus on key periods in the signal (such as beam state mutation points). This design not only improves the model's sensitivity to abnormal signals but also enhances the interpretability of the results \cite{vaswani2017}. Experiments show that this combined architecture can effectively handle the dynamic prediction of individual bunch parameters in storage rings.

\subsubsection{Dynamic Feature Fusion Module}
\label{sec:fusion}

The Dynamic Feature Fusion Module is an innovative component of this study, designed to achieve adaptive fusion of temporal features and auxiliary compensation features. This module employs a learnable dynamic weighting mechanism to optimally combine the temporal features (attended\_out) extracted from the LSTM+Attention path with the Tshift compensation information, significantly improving the accuracy of beam parameter prediction.

The module adopts a three-layer processing architecture: first, the Tshift compensation information undergoes a nonlinear transformation via a fully connected layer, mapping it to the same feature space as the main path features. The mathematical expression is as follows:

\begin{gather}
X_{Tshift} = \text{ReLU}(W_{e} \cdot Tshift + b_{e}) \label{eq:tshift_transform} \\
z = \alpha \cdot \text{attended}_{out} + (1 - \alpha) \cdot X_{Tshift} \label{eq:fusion}
\end{gather}

where $X_{Tshift}$ represents the transformed feature vector, $W_{e}$ is the weight matrix, $b_{e}$ is the bias vector, and ReLU($\cdot$) is the Rectified Linear Unit activation function. Subsequently, dynamic fusion weights $\alpha = \sigma(w)$ are generated through a trainable parameter $w$, where $\sigma(\cdot)$ denotes the sigmoid function, ensuring the weight range is (0,1). The final feature fusion is completed using the weighted sum shown in the following equation, where $z$ is the fused feature vector.

This design has three significant advantages: firstly, it automatically optimizes the fusion weights through end-to-end training, avoiding the limitations of manually setting fixed coefficients; secondly, it uses nonlinear transformation to ensure compatibility between auxiliary features and main path features; finally, it employs a dynamic adjustment mechanism to adapt to the feature fusion requirements under different beam states. Experiments show that this dynamic fusion strategy more effectively preserves the temporal characteristics of the main path features while fully utilizing the compensatory role of auxiliary information compared to traditional methods.

\subsection{Joint Model Establishment}
\label{sec:joint_model}

\begin{figure}[h]
\centering
\includegraphics[width=0.8\linewidth]{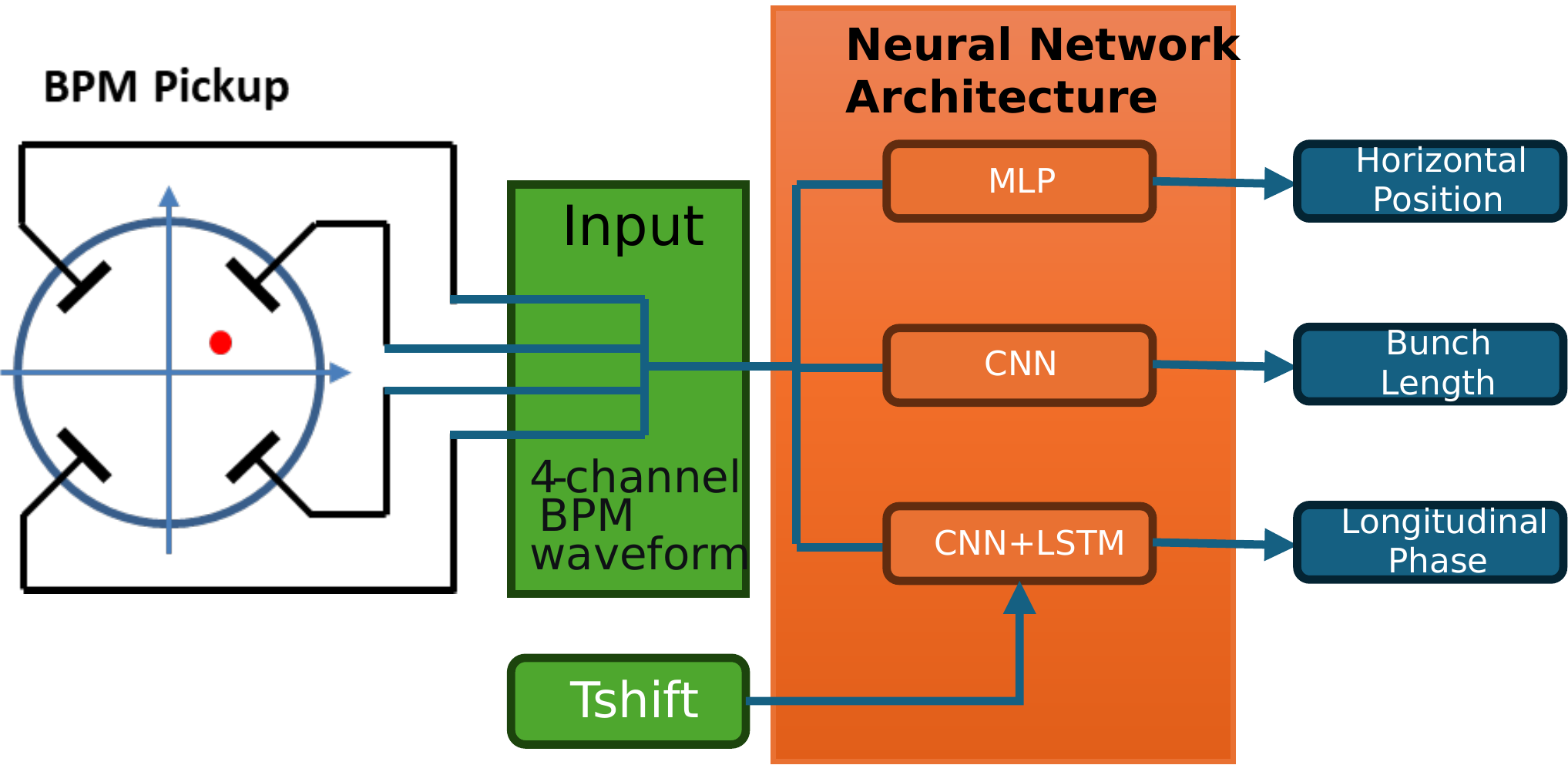}
\caption{Architecture of the joint prediction model for bunch-by-bunch beam parameter estimation.}
\label{fig:joint_model}
\end{figure}

The joint prediction model proposed in this study integrates the aforementioned specialized neural network modules to achieve collaborative prediction of transverse position, bunch length, and phase for individual bunches in the storage ring. As shown in Fig. \ref{fig:joint_model}, this joint model adopts a divide-and-conquer strategy, designing optimal sub-network structures for the characteristic differences of different physical quantities, and realizes end-to-end training and prediction through a unified input-output framework.

\section{Application in Dataset}
\label{sec:application}

Based on the joint prediction model framework established in Section \ref{sec:model}, this chapter details the specific application and implementation of each sub-module on the measured dataset from the Hefei Light Source. The Hefei Light Source (HLS-II), as an upgraded second-generation synchrotron radiation light source, features a beam diagnostic system that collects BPM signals with typical high-frequency sampling (78 sampling points per waveform) and multi-electrode synchronization (4 electrode channels), providing an ideal data foundation for validating model performance. This set of experimental data was collected during steady-state operation. All data underwent strict preprocessing, including noise filtering, time alignment, and amplitude normalization, to ensure the quality and consistency of the input data.

\subsection{Transverse Position Prediction}
\label{sec:position_prediction}

\begin{figure}[h]
\centering
\includegraphics[width=0.8\linewidth]{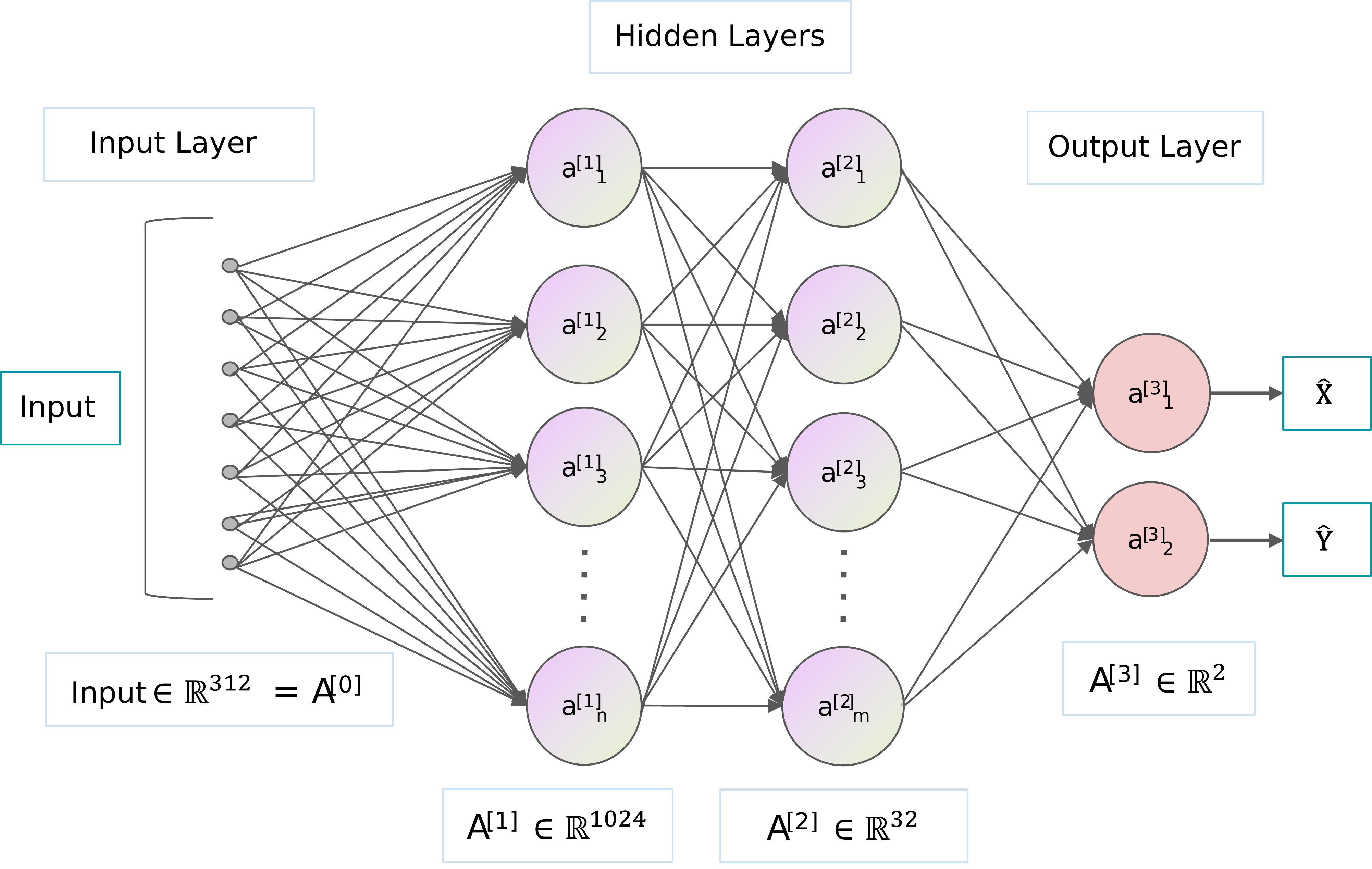}
\caption{MLP-based beam transverse position prediction model with two hidden layers.}
\label{fig:mlp_model}
\end{figure}

This study employs a Multilayer Perceptron (MLP) architecture to accomplish the prediction task of the beam's transverse position in the storage ring. The model takes the raw waveform signals from four BPM electrodes as input, with each electrode containing 78 time sampling points, resulting in an input data dimension of [4, 78]. The model structure is shown in Fig. \ref{fig:mlp_model}.

The model first flattens the multi-dimensional input into a one-dimensional vector, converting the $4\times78$ two-dimensional structure into a 312-dimensional feature representation. This flattening process allows the model to directly capture the global correlation characteristics between the amplitudes of the electrode signals, which aligns well with the difference-over-sum physical principle used in calculating the beam's transverse position.

This is followed by two hidden layers with batch normalization. The first fully connected layer maps the 312-dimensional input to a 1024-dimensional feature space, followed by batch normalization and ReLU activation function processing. This mapping is accomplished through a learnable weight matrix. Batch normalization accelerates model convergence by standardizing the input distribution of each layer. The second fully connected layer further compresses the feature dimension to 32 dimensions, maintaining the same batch normalization and activation processing flow.

The output layer uses a linear transformation to map the 32-dimensional features to a two-dimensional coordinate space, directly predicting the beam's (x, y) position information, i.e., the horizontal and vertical positions. This process is also completed through a learnable weight matrix. The ReLU activation function used in the model design introduces nonlinear capability by retaining positive values and suppressing negative values, enabling the model to learn complex features. Simultaneously, its gradient being constant at 1 in the positive region effectively alleviates the vanishing gradient problem in deep networks. This concise MLP structure has significant computational efficiency advantages, enabling millisecond-level real-time prediction in GPU environments.

\subsection{Bunch Length Prediction}
\label{sec:length_prediction}

This study constructs a bunch length prediction model based on a one-dimensional Convolutional Neural Network (1D-CNN). The model achieves efficient processing of multi-channel time-domain waveform signals from the HLS-II BPM through hierarchical feature extraction. The model adopts an input dimension of [4,78], where 4 represents the number of BPM channels and 78 is the number of time sampling points per channel, fully preserving the spatiotemporal characteristics of the original signal.

In the feature extraction stage, the model gradually abstracts signal features through two convolution-pooling modules. The first convolutional layer uses 16 kernels of width 3 for feature extraction. Here, the convolution operation maintains the same output temporal length as the input (78 points) by padding 1 zero on each end of the input sequence (padding=1). This process can be represented as:

\begin{equation}
\mathbf{f}_{1} = \text{ReLU}\left( \mathbf{W}_{1}*\mathbf{X} + \mathbf{b}_{1} \right) \in \mathbb{R}^{\text{batch} \times 16 \times 78}
\label{eq:conv1}
\end{equation}

where $\mathbf{W}_{1} \in \mathbb{R}^{16 \times 4 \times 3}$ represents the convolutional kernel weights, and $*$ denotes the 1D convolution operation.

This is followed by downsampling. The max-pooling layer uses a sliding window of width 2 (kernel\_size=2), selecting the maximum value within each window along the temporal dimension with a stride of 2. This operation compresses the temporal dimension to half its original size:

\begin{equation}
\mathbf{p}_{1} = \text{MaxPool1d}\left( \mathbf{f}_{1} \right) \in \mathbb{R}^{\text{batch} \times 16 \times 39}
\label{eq:pool1}
\end{equation}

The second convolutional layer uses 32 kernels of width 3, also maintaining the temporal length by padding 1 point:

\begin{equation}
\mathbf{f}_{2} = \text{ReLU}\left( \mathbf{W}_{2}*\mathbf{p}_{1} + \mathbf{b}_{2} \right) \in \mathbb{R}^{\text{batch} \times 32 \times 39}
\label{eq:conv2}
\end{equation}

After processing by a max-pooling layer with the same parameters, a final feature representation with dimensions [batch, 32, 19] is obtained. These alternating convolution and pooling operations effectively capture local fluctuations and global evolution patterns in the beam signal, retaining key temporal features while achieving efficient data compression.

In the feature mapping stage, the model flattens the feature tensor processed by convolution and pooling into a one-dimensional feature vector. This step reshapes the feature tensor of dimension [batch, 32, 19] into a vector form of [batch, 608], preparing it for subsequent fully connected layer processing. The flattened feature vector first passes through a first fully connected layer with 64 neurons, which also uses the ReLU activation function. Subsequently, the feature vector is passed to the final output layer, which maps the 64-dimensional features to a single bunch length prediction value. The entire feature mapping process achieves the transformation from a high-dimensional feature space to the prediction target while maintaining computational efficiency. Through this design, the model effectively correlates the extracted temporal features with the bunch length prediction task, thereby achieving accurate estimation of the bunch length parameter.

\begin{figure}[h]
\centering
\includegraphics[width=0.8\linewidth]{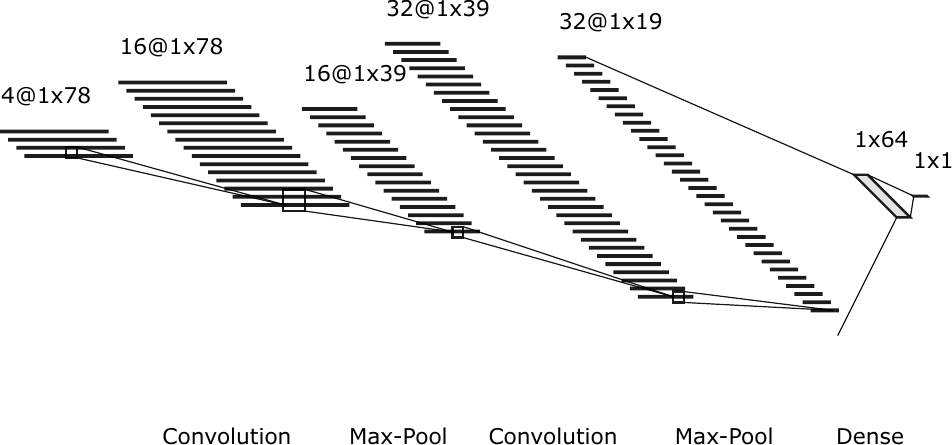}
\caption{1D-CNN based bunch length prediction model incorporating two convolutional, two pooling, and one fully-connected layer.}
\label{fig:cnn_model}
\end{figure}

The model achieves a dimensional transformation from [batch,4,78] to [batch,1] throughout the processing pipeline, with clear evolution of feature dimensions at each stage: the original input signal is converted to intermediate representations of [batch,16,78] and [batch,32,39] through convolutional layers, pooling operations gradually compress the temporal dimension to 19, and finally mapping from feature space to the predicted value is achieved through fully connected layers. This architectural design preserves the temporal characteristics of the signal while enabling end-to-end accurate prediction through deep feature extraction. The model structure is shown in Fig. \ref{fig:cnn_model}.

\subsection{Phase Prediction}
\label{sec:phase_prediction}

The phase prediction model proposed in this study adopts an innovative hybrid LSTM-CNN-Attention architecture, specifically designed to process temporal signal data from four Beam Position Monitors (BPM) and additional Tshift time compensation information. The core design of this model consists of three key components.

The first is the same feature extraction framework used in bunch length prediction, processing the raw signals from four BPM electrodes through a two-level 1D Convolutional Neural Network, ensuring consistency in the feature extraction method.

Secondly, the model introduces a Bidirectional LSTM network to process temporal features. This network contains two hidden layers, each with 64 units. Through its carefully designed gating mechanism, it can model long-range temporal dependencies in the BPM signals based on the previously extracted feature tensor of [batch, 32, 19]. The calculation processes for the various gating units are as follows \cite{greff2016}. The input gate $\mathbf{i}_{t}$ controls the update of new information:

\begin{equation}
\mathbf{i}_{t} = \sigma\left( \mathbf{W}_{i}\left[ \mathbf{h}_{t - 1},\mathbf{x}_{t} \right] + \mathbf{b}_{i} \right)
\label{eq:input_gate}
\end{equation}

where $\mathbf{h}_{t - 1}$ represents the hidden state from the previous time step, $\mathbf{x}_{t}$ represents the input feature at the current time step, $\mathbf{W}_{i}$ is the input gate weight matrix, $\mathbf{b}_{i}$ is the bias term, and $\sigma$ is the sigmoid function.

The forget gate determines the degree of retention of historical information:

\begin{equation}
\mathbf{f}_{t} = \sigma\left( \mathbf{W}_{f}\left[ \mathbf{h}_{t - 1},\mathbf{x}_{t} \right] + \mathbf{b}_{f} \right)
\label{eq:forget_gate}
\end{equation}

The output gate regulates the output of the hidden state:

\begin{equation}
\mathbf{o}_{t} = \sigma\left( \mathbf{W}_{o}\left[ \mathbf{h}_{t - 1},\mathbf{x}_{t} \right] + \mathbf{b}_{o} \right)
\label{eq:output_gate}
\end{equation}

The cell state is the core component of the LSTM, maintaining a linear flow throughout the time sequence propagation, selectively updating or forgetting information through the gating mechanism, thus achieving long-term memory function. Its update process is as follows:

\begin{equation}
\tilde{\mathbf{C}}_{t} = \tanh\left( \mathbf{W}_{C}\left[ \mathbf{h}_{t - 1},\mathbf{x}_{t} \right] + \mathbf{b}_{C} \right)
\label{eq:candidate_cell}
\end{equation}

\begin{equation}
\mathbf{C}_{t} = \mathbf{f}_{t} \odot \mathbf{C}_{t - 1} + \mathbf{i}_{t} \odot \tilde{\mathbf{C}}_{t}
\label{eq:cell_state}
\end{equation}

$\tilde{\mathbf{C}}_{t}$ is the candidate memory content at the current time step, $\mathbf{C}_{t - 1}$ is the cell state from the previous time step, and $\odot$ denotes element-wise multiplication.

The final hidden state output:

\begin{equation}
\mathbf{h}_{t} = \mathbf{o}_{t} \odot \tanh\left( \mathbf{C}_{t} \right)
\label{eq:hidden_state}
\end{equation}

\begin{figure}[h]
\centering
\includegraphics[width=0.8\linewidth]{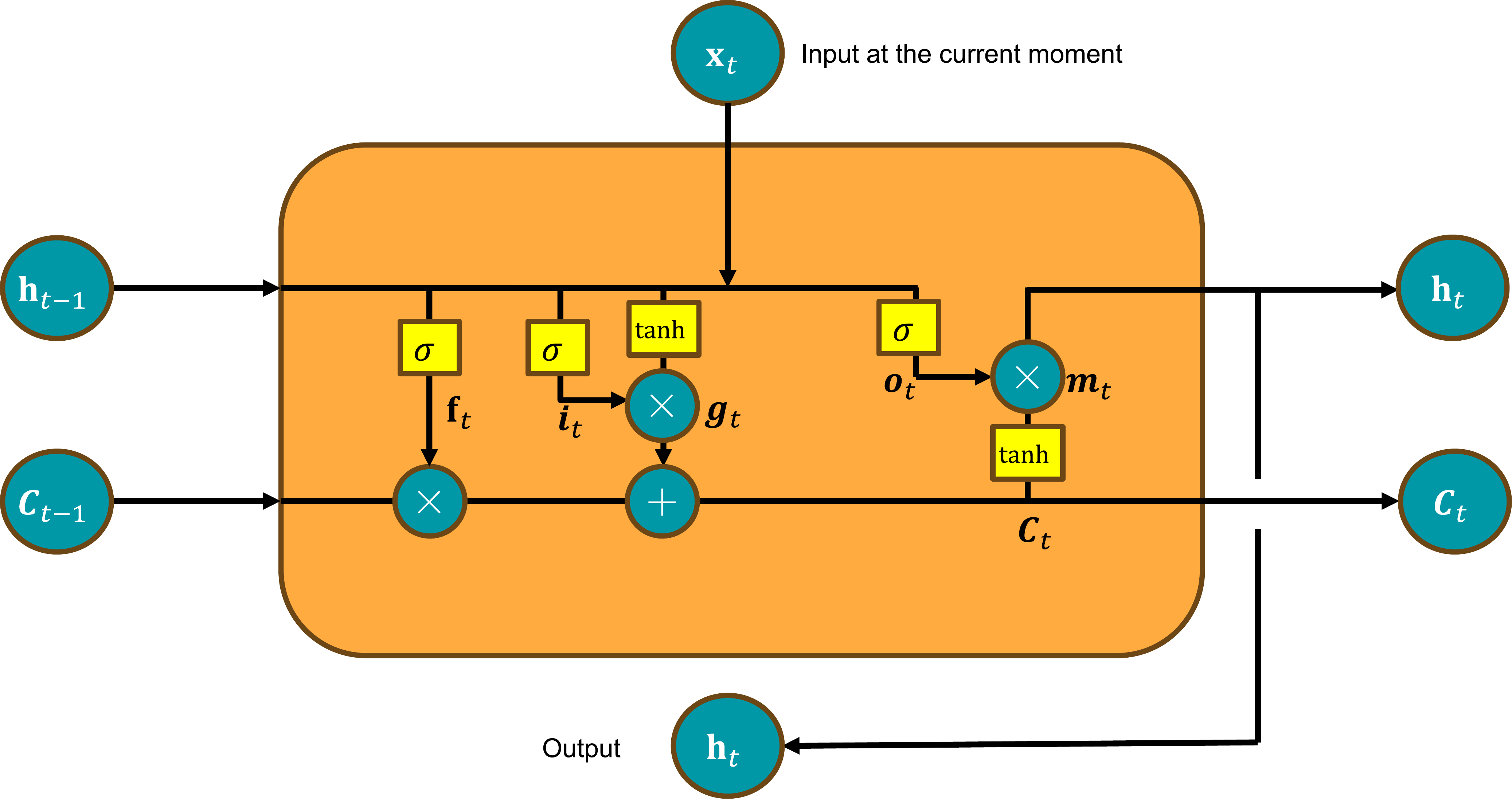}
\caption{Operating principle of the LSTM stage.}
\label{fig:lstm}
\end{figure}

Specifically, we added an attention mechanism to the LSTM output. This part receives the feature data output by the LSTM layer with shape [batch\_size, 19, 64]. It first calculates the importance weights for each time step through two fully connected layers. The weight calculation process uses parallelized matrix operations instead of loop processing to improve computational efficiency. After softmax normalization, an attention weight matrix with shape [batch\_size, 19, 1] is obtained. Finally, the model uses these weights to perform a weighted sum on the LSTM outputs, generating a context vector with dimensions [batch\_size, 64].

In the feature fusion stage, the model innovatively introduces a dynamic weighting strategy, namely the Dynamic Feature Fusion Module introduced in the previous section. The time compensation parameter Tshift, after processing by a fully connected layer, is also mapped to a vector with shape [batch\_size, 64]. It is then adaptively fused with the attention-weighted context vector. The fusion weight is controlled by a learnable parameter, using the sigmoid function to ensure the weight value smoothly varies within the range (0,1). This design preserves the physical significance of traditional compensation methods while empowering the model with the ability to autonomously adjust feature importance.

The entire model's data flow undergoes a clear dimensional evolution: from the initial input of [batch\_size, 4, 78], to [batch\_size, 32, 19] after CNN processing, converted to [batch\_size, 19, 64] by the LSTM, and finally outputting a prediction result of [batch\_size, 1] through the attention mechanism and feature fusion.

This architecture has three main advantages: firstly, the combination of CNN and LSTM can simultaneously capture local and global temporal features of the signal; secondly, the attention mechanism can automatically focus on key time regions of the signal; thirdly, the dynamic fusion strategy balances data-driven features and traditional compensation methods. Experiments show that this design significantly improves the accuracy and robustness of phase prediction while maintaining physical plausibility.

\section{Result Analysis}
\label{sec:results}

\begin{figure}[h]
\centering
\includegraphics[width=0.8\linewidth]{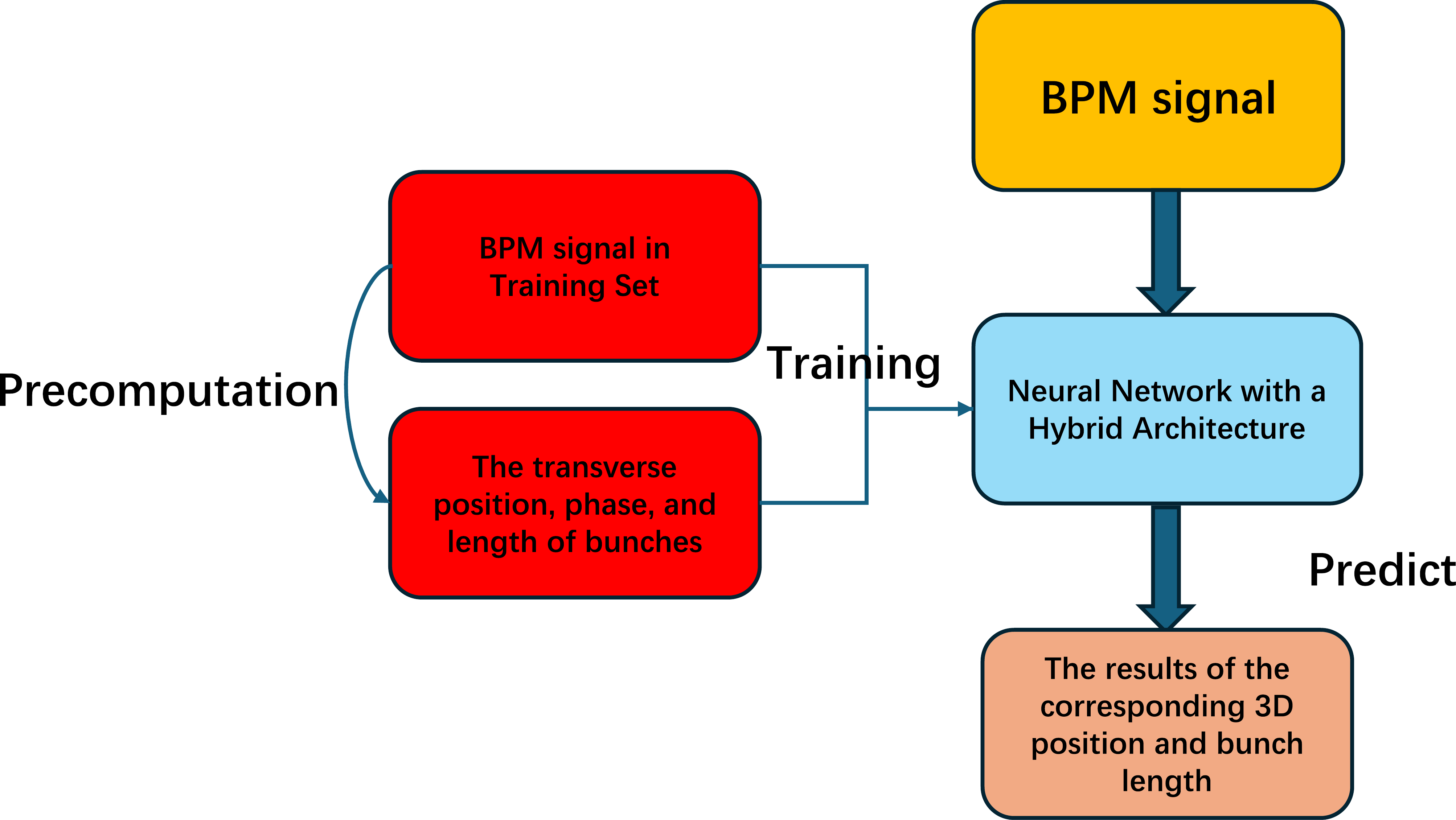}
\caption{Machine learning schematic for bunch-by-bunch prediction of 3D position and length.}
\label{fig:workflow}
\end{figure}

This study was conducted in an accelerated computing environment based on an NVIDIA RTX 4090 graphics card. The Adam optimizer (initial learning rate 0.005) and a StepLR learning rate scheduling strategy (decaying to 0.98 times the current rate every 10 epochs) were used. All models were fully trained for 2000 epochs. The Mean Squared Error (MSE) was used as the loss function during training, and the batch size was set to 256 to ensure stable optimization of model parameters. The training datasets came from two parts: Shanghai Light Source and Hefei Light Source. Two separate models were trained from these respective datasets. The analysis below also pertains to the prediction results of each model on its respective test set. The schematic diagram of the entire work is shown in Fig. \ref{fig:workflow}.

\subsection{Transverse Position Prediction Results}
\label{sec:position_results}

The transverse position prediction model was evaluated on both facilities, demonstrating robust performance. The coefficient of determination ($R^2$) and mean absolute error (MAE) are used as the primary evaluation metrics. $R^2$ measures the proportion of variance in the target data that is predictable from the input features, with a value of 1 indicating perfect prediction. MAE provides an intuitive measure of the average magnitude of prediction errors in the original units (mm), offering a clear interpretation of the model's accuracy.

\begin{figure}[h]
\centering
\includegraphics[width=0.8\linewidth]{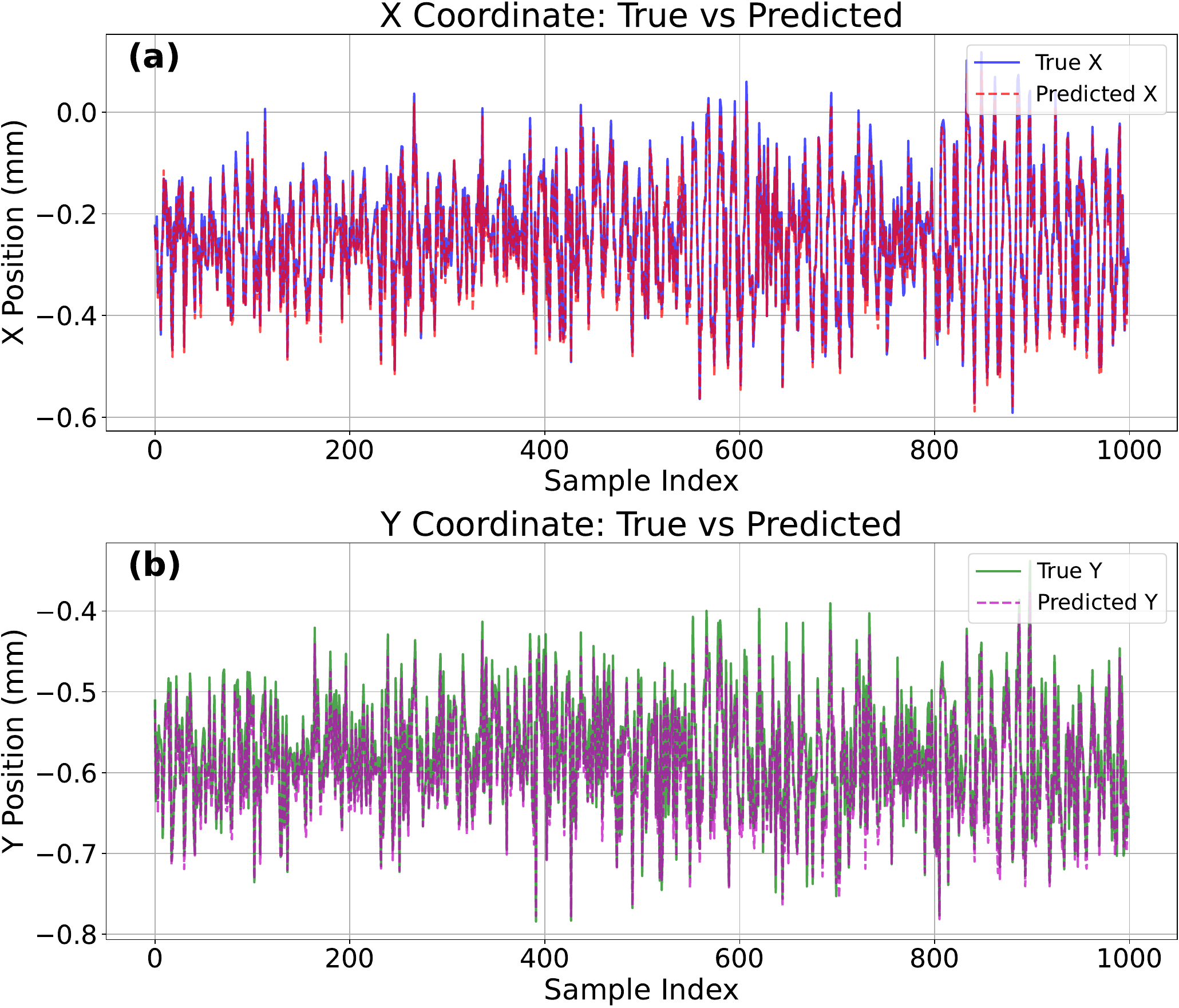}
\caption{Transverse position prediction results for HLS-II during steady-state operation. (a) Horizontal position (X). (b) Vertical position (Y). The solid and dashed lines represent the ground truth and model predictions, respectively, for a consecutive segment of 1,000 samples after sparse sampling.}
\label{fig:hls_position}
\end{figure}

For HLS-II, the model was tested on a large-scale dataset containing 99,178 samples, achieving outstanding performance with an $R^2$ of 0.991 and MAE of 0.0095 mm for the horizontal (X) position, and an $R^2$ of 0.986 and MAE of 0.0066 mm for the vertical (Y) position. The total inference time for this substantial dataset was 1.873 seconds. Fig. \ref{fig:hls_position} displays a representative segment of 1,000 consecutive samples from the steady-state operation, where subfigure (a) and (b) show the excellent agreement between the ground truth (solid lines) and predictions (dashed lines) for the X and Y positions, respectively. The high $R^2$ values and low MAEs confirm the model's precision in tracking both transverse dimensions under steady-state conditions.

\begin{figure}[h]
\centering
\includegraphics[width=0.8\linewidth]{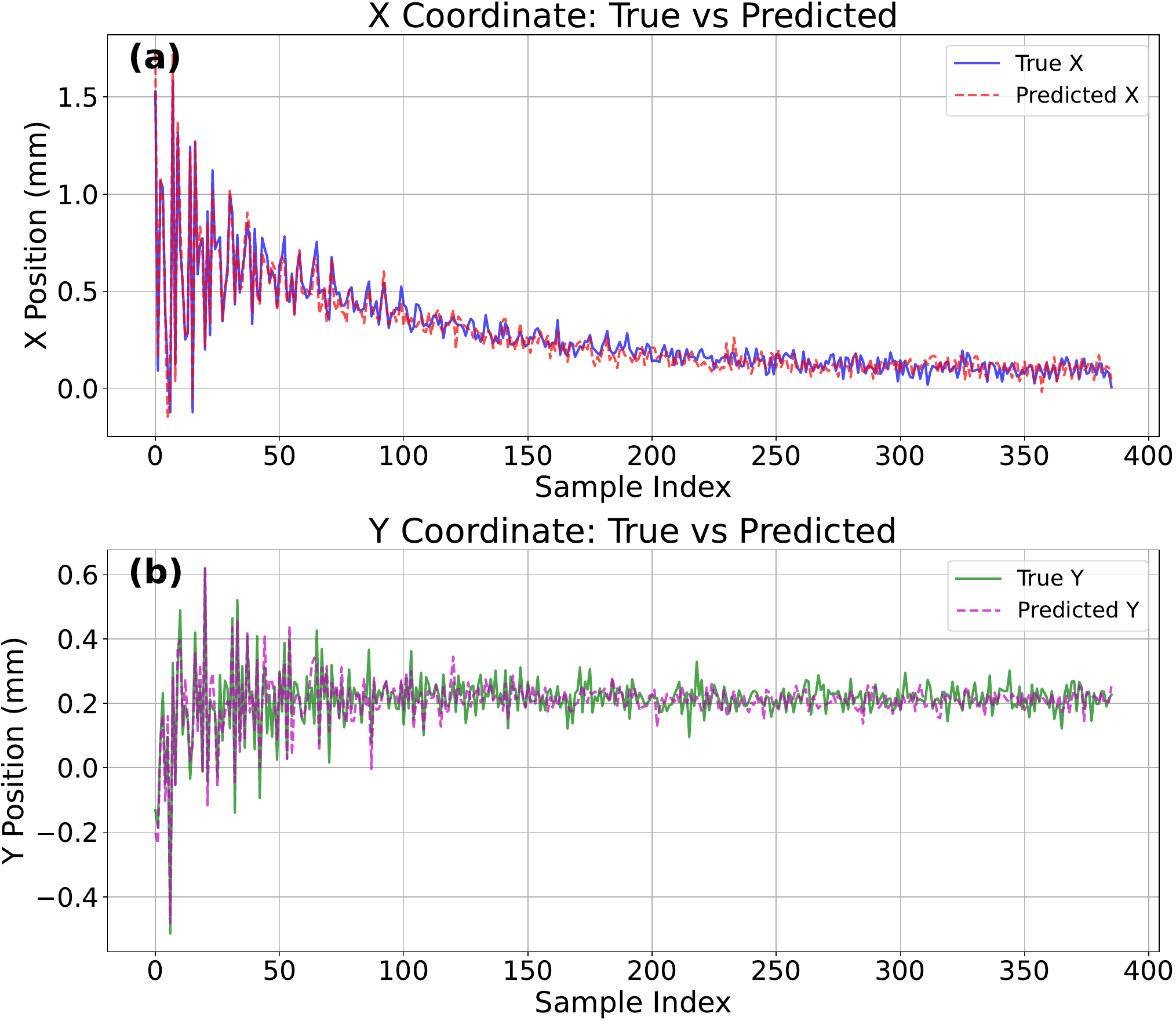}
\caption{Transverse position prediction results for SSRF Bunch 5. The plots display the evolution from injection transients to steady state. (a) Horizontal position (X). (b) Vertical position (Y). In both subfigures, the solid lines represent the ground truth while the dashed lines represent the model predictions.}
\label{fig:ssrf_position}
\end{figure}

In the case of SSRF data, the prediction encompassed 387 bunch samples, covering the complete evolution of Bunch 5 from the test set through injection transients to steady-state stabilization. The total inference time for this dataset was 0.033 seconds. As illustrated in Fig. \ref{fig:ssrf_position}, the model showed differing capabilities in tracking horizontal and vertical motions. For the horizontal direction (subfigure a), predictions closely followed the actual trajectory with an $R^2$ of 0.949 and MAE of 0.043mm. Vertical position tracking (subfigure b) proved more challenging, though the model still captured the general trend with $R^2$ of 0.726 and MAE of 0.036mm, particularly during the initial injection oscillations where signal features were more pronounced.

A common characteristic observed in both SSRF and HLS-II predictions is that the most pronounced errors tend to occur at the peaks and troughs of the position trajectories. This is largely because these extreme points inherently occur less frequently in the dataset, resulting in insufficient training samples for the model to learn from effectively.

However, a fundamental difference in performance is evident. The model achieves remarkably high accuracy on the HLS-II dataset, which can be attributed to its steady-state operation and the resulting high signal-to-noise ratio. Furthermore, the HLS-II data contains 78 sampling points per waveform, and this higher data density provides the model with richer feature information. The stable beam conditions provide a clean and consistent mapping from BPM signals to beam positions, which the model can learn effectively.

In contrast, the prediction for SSRF data, especially for the vertical position as it flattens, shows significant errors. A key observation is that the fluctuation range of the predicted values is noticeably smaller than that of the ground truth. This "underestimation" of dynamics points to a potential information loss or feature smoothing within the model when processing SSRF signals. An important technical factor behind this is the relatively insufficient sampling rate of the SSRF data, with only 32 sampling points per waveform, which limits the model's ability to extract subtle features from the signals. The underlying cause may also be linked to the injection transients present in the SSRF data. The model, trained on data containing both large injection oscillations and small steady-state fluctuations, might have learned a compromised representation that fails to resolve the subtle features indicative of small-amplitude motions in the steady state as effectively. Furthermore, differences in BPM response characteristics or beam coupling impedances between the two rings could make the extraction of vertical position from the electrode signals inherently more challenging for SSRF in this specific regime.

\subsection{Bunch Length Prediction Results}
\label{sec:length_results}

The performance of the bunch length prediction model was evaluated on the HLS-II dataset. The model demonstrated exceptional accuracy in estimating bunch length from the raw BPM waveforms. On the full test set comprising 99,178 samples, the model achieved a near-perfect coefficient of determination ($R^2$) of 0.997, with a mean absolute error (MAE) of 1.136 femtoseconds (fs). The total inference time for the entire dataset was 1.04 seconds, underscoring the model's high computational efficiency and suitability for real-time applications.

\begin{figure}[h]
\centering
\includegraphics[width=0.8\linewidth]{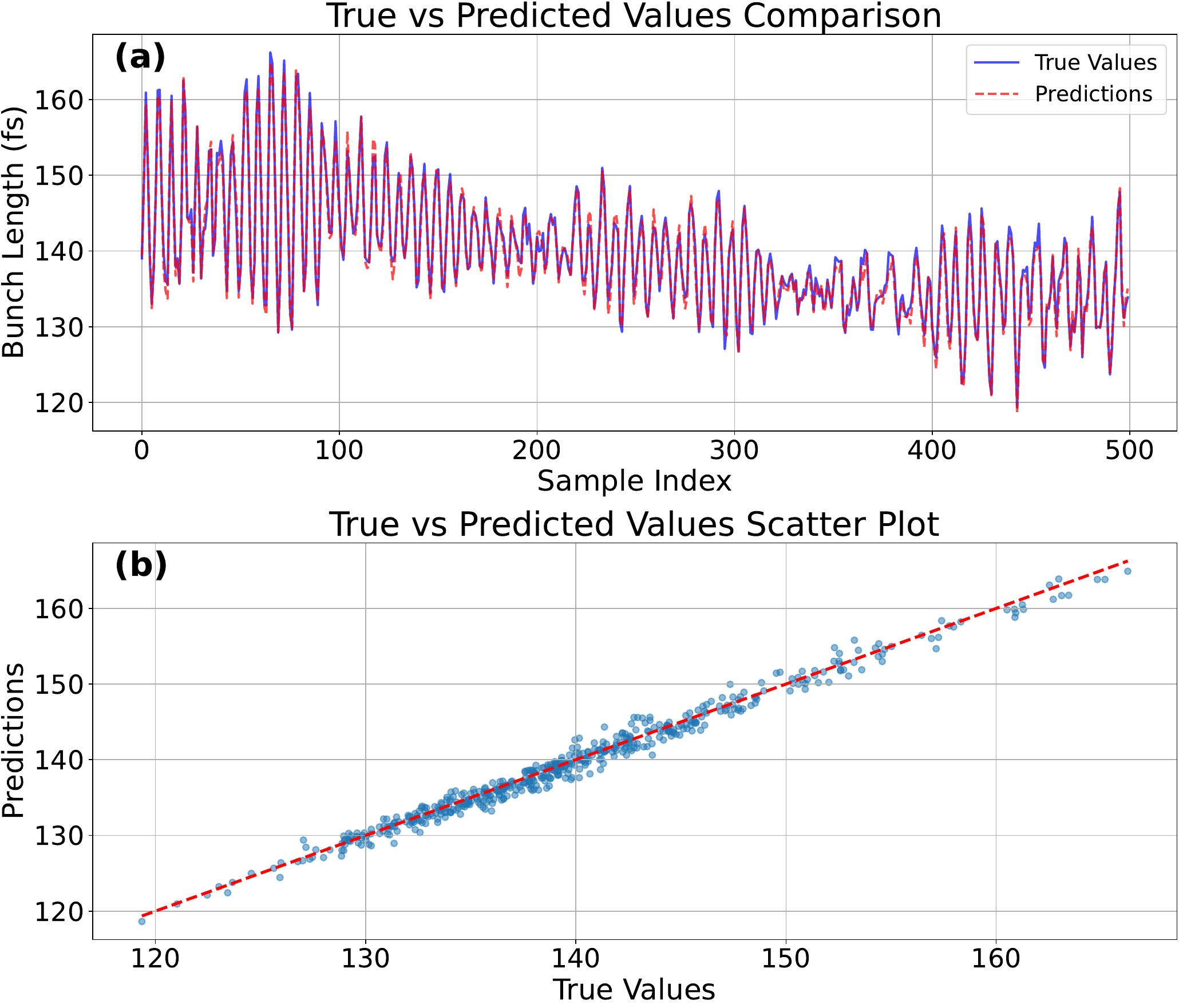}
\caption{Performance evaluation of the HLS-II bunch length prediction model: (a) temporal prediction vs. ground truth, (b) prediction scatter plot.}
\label{fig:bunch_length_results}
\end{figure}

Fig. \ref{fig:bunch_length_results} presents the performance evaluation results of the HLS-II bunch length prediction model. Subfigure (a) displays the prediction effectiveness for 500 consecutive samples selected via sparse sampling from the steady-state operation data, comparing the temporal evolution of the ground truth bunch length (solid line) against the model predictions (dashed line). The close alignment between the two curves demonstrates the model's remarkable capability to accurately reconstruct the periodic variations in bunch length over time. Subfigure (b) shows the scatter plot of predicted versus true values across the entire test set. The dense clustering of data points along the diagonal line $y = x$ further confirms the model's outstanding prediction accuracy across the full range of bunch lengths present in the data.

\subsection{Longitudinal Phase Prediction Results}
\label{sec:phase_results}

The phase prediction model demonstrated high accuracy across both facilities, effectively capturing the longitudinal phase dynamics from BPM waveforms.

\begin{figure}[h]
\centering
\includegraphics[width=0.8\linewidth]{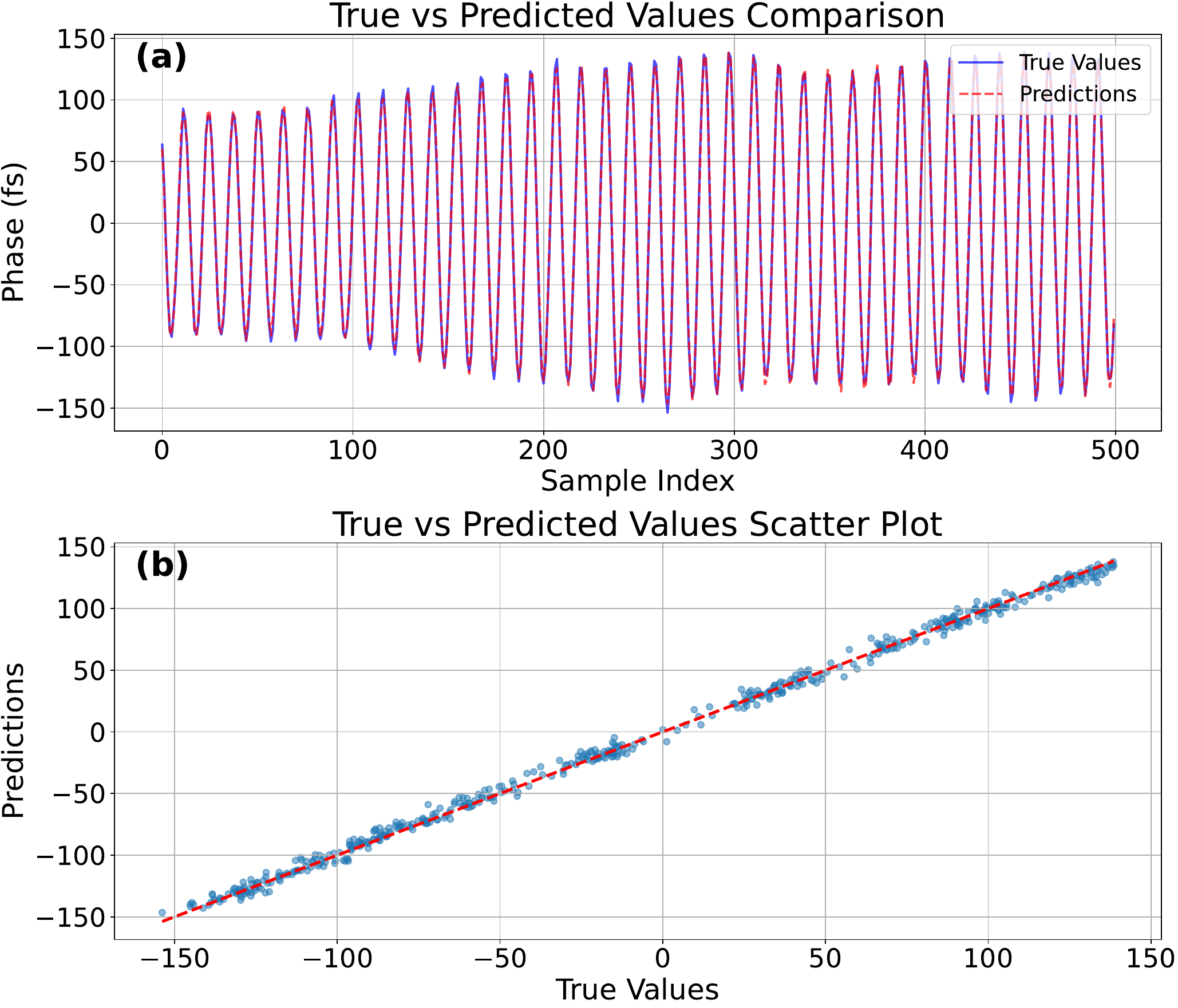}
\caption{Performance evaluation of the HLS-II longitudinal phase prediction model.}
\label{fig:hls_phase}
\end{figure}

On the HLS-II dataset, the model demonstrated high predictive accuracy, achieving an $R^{2} = 0.973$ and a Mean Absolute Error (MAE) of 3.59 fs, with a total inference time of 1.19 seconds. The prediction results for a representative subset of samples are visualized in Fig. \ref{fig:hls_phase}. As illustrated in subfigure (a), the "True vs. Predicted Values Comparison" shows a blue solid line representing the ground truth and a red dashed line representing the model's predictions. The two curves exhibit close alignment across the 500-sample index, indicating the model's capacity to track the rapid oscillations in phase. This performance is further validated by the "True vs. Predicted Values Scatter Plot" in subfigure (b), where the data points are tightly clustered along the identity line. The residuals are uniformly distributed across the operational range of -150 fs to +150 fs, demonstrating consistent predictive reliability and a lack of significant bias across the entire phase interval.

\begin{figure}[h]
\centering
\includegraphics[width=0.8\linewidth]{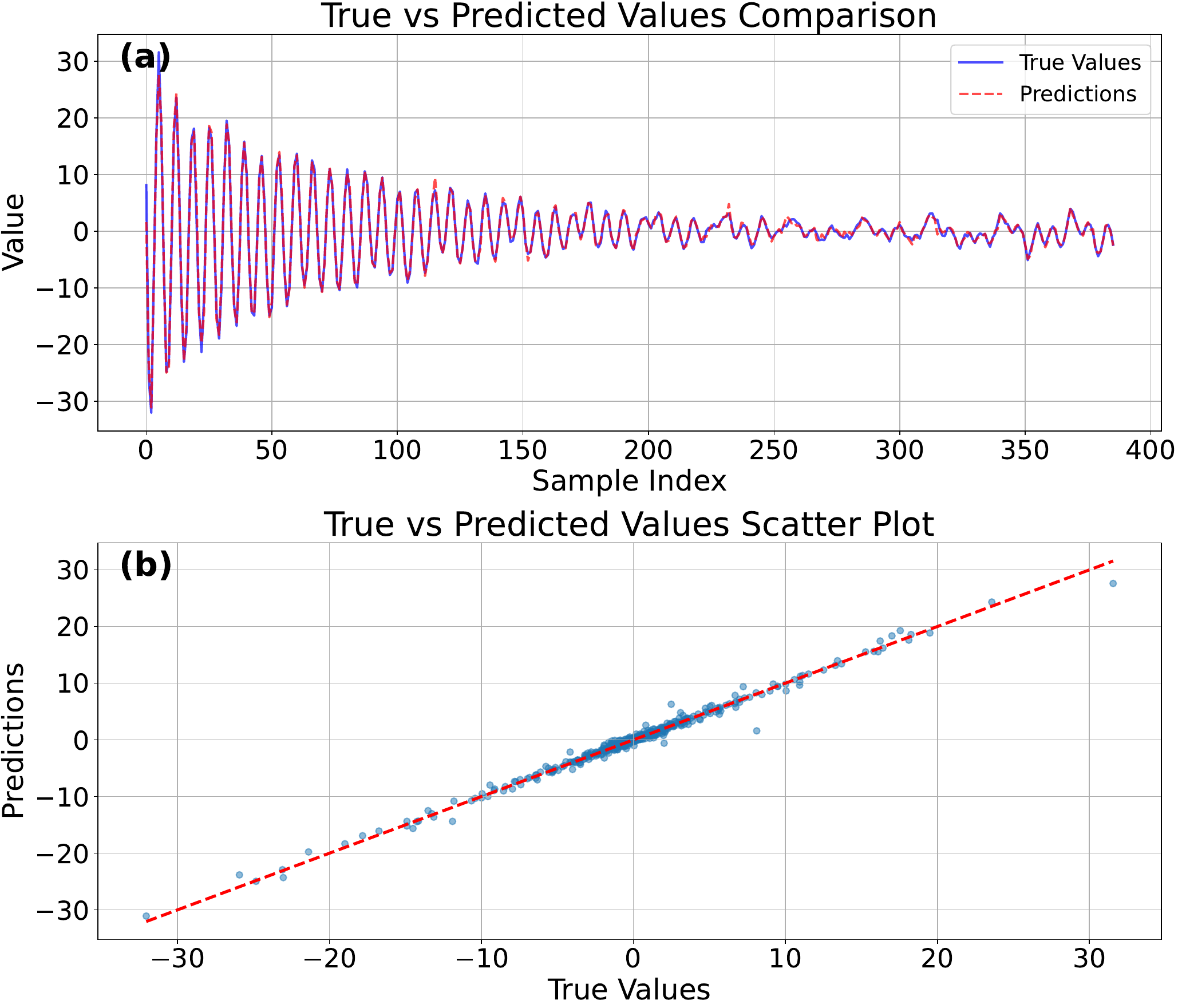}
\caption{Performance evaluation of the SSRF longitudinal phase prediction model.}
\label{fig:ssrf_phase}
\end{figure}

The model also demonstrated exceptional performance on the SSRF dataset, achieving an $R^2$ of 0.9886 and an MAE of 0.5040 fs across 387 samples, with a total inference time of 0.10 seconds. The prediction results for SSRF are shown in Fig. \ref{fig:ssrf_phase}, illustrating high-fidelity tracking of the phase evolution. Although a marginal increase in prediction residuals is observed near the phase extrema and zero-crossings, the model maintains high accuracy across the full signal profile. This slight deviation at turning points is likely attributable to the heightened nonlinearity in the mapping relationship between the beam position monitor (BPM) waveform and the beam phase in these specific intervals.

The model’s consistent high performance across two distinct experimental facilities confirms the strong generalizability and robustness of the proposed phase prediction architecture. Such high-fidelity phase prediction enables its direct application in advanced beam dynamics studies, such as the experimental analysis of beam loading effects in storage rings \cite{zhou2023}.

\section{Conclusion}
\label{sec:conclusion}

Addressing the pressing need for real-time, multi-parameter diagnostics in diffraction-limited storage rings (DLSRs), this study successfully developed an innovative neural network-based framework for bunch-by-bunch beam diagnostics. The framework simultaneously predicts the key beam parameters of transverse position, longitudinal phase, and bunch length based on raw beam position monitor (BPM) electrode signals. To accommodate the distinct physical characteristics of each parameter, the framework employs specialized subnetworks within a hybrid architecture that integrates a computationally efficient multilayer perceptron, long short-term memory networks, convolutional layers, and attention mechanisms.

The model was rigorously validated using experimental data from both the Shanghai Synchrotron Radiation Facility and the Hefei Light Source, demonstrating excellent accuracy and robustness. For the steady-state operating data from the Hefei Light Source, the predictions of all three parameters achieved R² values above 0.97. On the more complex Shanghai data, which include transient injection processes, the predictions of longitudinal phase and horizontal position similarly showed outstanding performance. Although vertical position prediction remains challenging in certain scenarios, the overall results confirm the effectiveness and applicability of the proposed framework.

This approach overcomes the limitations of traditional algorithms by eliminating the constraints of serial processing chains and batch-processing modes inherent in systems such as HOTCAP. By requiring only a short signal segment from a single bunch to deliver real-time multi-parameter predictions, it enables efficient processing without the need to pre-fit global response functions. Experimental results show that the theoretical latency for jointly predicting the three parameters of a single bunch is only 0.042 ms, laying the technical foundation for truly real-time, bunch-by-bunch diagnostics and closed-loop feedback.
Furthermore, the framework reveals intrinsic beam dynamics correlations that are often overlooked by conventional methods. Traditional approaches typically treat each bunch and each turn as independent measurements, whereas our method, through temporal modeling and attention mechanisms, inherently captures complex inter-bunch and turn-by-turn dependencies. While the current implementation processes single-turn data independently, future extensions could incorporate multi-turn sequences and multi-bunch interaction modeling, further enhancing prediction accuracy, enabling early detection of anomalous beam behavior, and supporting adaptive, predictive beam control strategies.

Beyond this study, this data-driven methodology offers significant extensibility. On one hand, it can be embedded into real-time feedback systems to actively suppress beam instabilities; on the other, it can be expanded to predict additional key parameters and transferred to next-generation facilities such as diffraction-limited storage rings, opening new avenues for intelligent beam operation and optimization.

\begin{acknowledgments}
This work was supported by the Major Science and Technology Infrastructure Maintenance and Renovation Project of the Chinese Academy of Sciences (Project: "General Signal Processing Platform for Accelerator Beam Diagnostics and Control"). The authors would like to thank the SSRF and HLS-II operations teams for their assistance in data acquisition.
\end{acknowledgments}

\bibliography{./apssamp}
\end{document}